\documentstyle [11pt]{article}

\begin{document}

\begin{center}
{\Large\bf
    INHOMOGENEOUS BOSON REALIZATIONS OF THE POLYNOMIAL
ANGULAR MOMENTUM ALGEBRA OF ARBITRARY POWER }
\end{center}

\vspace{3mm}

\begin{center}
DONG RUAN \footnote{E-mail: dongruan@tsinghua.edu.cn},
YUFENG JIA and HONGZHOU SUN\\
{\small\it 
	Department of Physics, Tsinghua University, Beijing 100084, 
P. R. China, 
	The Key Laboratory of Quantum Information and
Measurements of Ministry of Education, Tsinghua University,
Beijing 100084, P. R. China and 
	Center of Theoretical Nuclear Physics,
National Laboratory of Heavy Ion Accelerator,
Lanzhou, 730000, P. R. China
}
\end{center}

\vspace{1cm}

\begin{abstract}
The inhomogeneous single-, two- and three-boson realizations of the more 
general polynomial angular momentum algebra ${\cal SU}_n(2)$ are obtained 
from the Fock representations of ${\cal SU}_n(2)$ that corresponds to 
the indecomposable representation on the space of universal enveloping 
algebra $U({\cal SU}_n(2))$ and to the induced representations 
on the quotient spaces $U({\cal SU}_n(2))/I_i$, with $I_i$ being some
left ideals of $U({\cal SU}_n(2))$.
\end{abstract}

\newpage

Recently, many attention has been paid to a type of polynomial angular 
momentum algebras, denoted by ${\cal SU}_n(2)$ (the suffix $n$ is 
referred to as the highest power of polynomial), and their physical 
applications in quantum mechanics \cite{bht} such as additional symmetries, 
exactly solvable and (super)integrable problems, inverse scattering and 
conformal field theories, \cite{ssn,ags} quantum optics, \cite{kk,sbjps} 
and so on. 
Similar to the ordinary angular momentum algebra SU(2), \cite{bl}
${\cal SU}_n(2)$ is spanned by three elements \{$J_3$, $J_+$, $J_-$\}, 
and however, they satisfy the following commutation relations
\begin{equation}
    [ J_3, \hspace{1mm} J_{\pm} ] = \pm J_{\pm} ,  \hspace{4mm}
    [ J_+, \hspace{1mm} J_-] = \sum_{i=0}^{n} C_i J_3^i,
\label{n-am}
\end{equation}
where $C_i$'s are arbitrary real numbers. It is obvious from 
Eq. (\ref{n-am}) that ${\cal SU}_n(2)$ has a coset structure 
${\cal SU}_n(2) = h+ v$, \cite{rocek} i.e., 
the element $J_3$ of ${\cal SU}_n(2)$ forms the Lie algebra $h$=U(1); 
the remaining two elements $J_+$, $J_-$ $\in v$ transform according to
a representation of U(1), moreover here, different from SU(2), 
their commutator yields a polynomial function of $J_3 \in$ U(1) with 
arbitrary powers and coefficients. 
When $C_1=2$ (or $-2$), $C_0=C_j=0$ ($j>1$), ${\cal SU}_n(2)$ goes back to
SU(2) (or its non-compact type SU(1,1)). 
If the polynomial function takes $(q^{J_3} - q^{ - J_3})/ (q - q^{ -1})$
with a positive real number $q$, an infinite series involving all odd 
powers of $J_3$, then ${\cal SU}_n(2)$ becomes the well known quantum group
SU$_q(2)$. \cite{bh}
In fact, the first special case of polynomial angular momentum algebra is
the so-called Higgs algebra, its polynomial function being 
$C_1 J_3 + C_3 J_3^3$, which was used by Higgs \cite{higgs} 
to establish the existence of additional symmetries for the isotropic 
oscillator and Kepler potentials in a two-dimensional curved space. 
Later many authors studied in detail the polynomial angular momentum 
algebras of different powers and their extensions [7, 10-21]. 
After Biedenharn realized SU$_q(2)$ by introducing $q$-deformed boson,
\cite{bh} Daskaloyannis \cite{dask} and Bonatsos \cite{bkd}
discussed the polynomial angular momentum algebras by means of generalized 
deformed oscillator respectively, and Quesne \cite{quesne1} related them to 
generalized deformed parafermion in the study of the spectra of Morse
and modied P\"{o}schl-Teller potentials. 
Beckers {\it et al.} \cite{bbd} discussed 
single-variable differential realizations of ${\cal SU}_2(2)$ and the Higgs 
algebra, and gave a unitrary two-boson realization of the Higgs algebra. 
More recently, Ruan {\it et al.} \cite{rjr} obtained inhomogeneous boson and 
differential realizations of ${\cal SU}_2(2)$ from its indecomposable representations. 
In the present work we shall further study the inhomogeneous boson realizations
of the more general polynomial angular momentum algebra ${\cal SU}_n(2)$
characterized by Eq. (\ref{n-am}) for the arbitrary power $n$.

\par
Let us begin with calculating the indecomposable representations
of ${\cal SU}_n(2)$. In terms of the Poincar\'e-Birkhoff-Witt
theorem, \cite{jacobson,dixmier} a basis for universal
enveloping algebra $U({\cal SU}_n(2))$ of ${\cal SU}_n(2)$ can be
chosen as the following set of ordered elements
\begin{equation}
    \{ X(m_1,m_2,m_3) = J_+^{m_1} J_-^{m_2} J_3^{m_3} |
        m_1,m_2,m_3 \in \aleph \}, 
\label{b}
\end{equation}
$X(0,0,0)= 1 $ denotes the identity operator, where the symbol $\aleph$
denotes the set of non-negative integers. It follows that by acting with 
three elements $J_3$, $J_+$, $J_-$ from the left upon the basis (\ref{b})
we may obtain a master representation of ${\cal SU}_n(2)$
\begin{equation}
\begin{array}{rl}
    \rho (J_3) X(m_1,m_2,m_3) = & X(m_1,m_2,m_3+1)
            + (m_1-m_2) X(m_1,m_2,m_3), \\
    \rho (J_+) X(m_1,m_2,m_3) = & X(m+1,m_2,m_3), \\
    \rho (J_-) X(m_1,m_2,m_3) = & X(m_1,m_2+1,m_3)
        -\sum\limits_{h=0}^{n}
        \sum\limits_{i=h}^{n} \sum\limits_{j=0}^{m_1-1}
         C_i C_{i}^h  \\
    {} & \times (j-m_2)^{i-h} X(m_1-1,m_2,m_3+h),
\label{indecom-rep}
\end{array}
\end{equation}
where the symbol $C_{i}^h$ is the usual binomial coefficient
$C_{i}^h = i!/ h!(i - h)!$. In order to obtain the third equation
of Eq. (\ref{indecom-rep}), we have used the following equations
with respect to Eq. (\ref{n-am})
\begin{equation}
\begin{array}{rl}
    [(J_3+j)^l, \hspace{1mm} J_\pm^m] = & \sum\limits_{i=0}^{l-1} C_l^i
        (\pm m)^{l-i} J_{\pm}^m (J_3+j)^i,     \hspace{3mm}  j\in \aleph,   \cr
    [J_-, \hspace{1mm} J_+^m]= & - J_+^{m-1}
    \sum\limits_{i=0}^{n} \sum\limits_{k=0}^{m-1}
    C_i (J_3+ k)^i.
\end{array}
\end{equation}
When $C_i=0$ ($i>2$), Eq. (\ref{indecom-rep}) leads to the results of 
${\cal SU}_2(2)$ obtained in Ref. \cite{rjr}.

We notice that the representation $\rho$ given by Eq. (\ref{indecom-rep}) 
in fact is indecomposable \cite{br,han} in $m_2$ and $m_3$, owing to the fact
that the values for the indices $m_2$ and $m_3$ do not decrease under
the action of $\rho$. Furthermore, on the quotient space 
$U({\cal SU}_n(2))/I_i$, where $I_i$'s are left ideals with respect to 
$U({\cal SU}_n(2))$, the indecomposable representation $\rho$ may induce the other
representations of ${\cal SU}_n(2)$ by choosing different $I_i$'s.

\par
Now let us construct the boson realizations of ${\cal SU}_n(2)$
from the indecomposable representation $\rho$. First of all, it can be seen
from Eq. (\ref{indecom-rep}) that the matrix elements
$\rho(J_{\alpha})_{m_1,m_2,m_3}^{m_1',m_2',m_3'}$ ($\alpha=3$, $+$, $-$), 
which are determined by the commutation relations (\ref{n-am}), are related to
three independent parameters $m_1$, $m_2$, $m_3$, so we need three sets
of independent boson operators $a_i^+$, $a_i$ ($i=1$, 2, 3) to
define a Fock space, which is automorphic to $U({\cal SU}_n(2))$, with basis 
\begin{equation}
    \{ |m_1,m_2,m_3 \rangle = (a_1^+)^{m_1} (a_2^+)^{m_2}
    (a_3^+)^{m_3} | 0 \rangle \hspace{1mm} |
    \hspace{1mm} m_1,m_2,m_3 \in \aleph \},
\label{fock-space}
\end{equation}
where $| 0 \rangle$ stands for a vacuum state, and $a_i |0 \rangle = 0$.
These boson operators, together with particle number operators
$\hat{n}_i$ $\equiv$ $a^+_i a_i$ ($i$=1, 2, 3), satisfy the commutation relations
\begin{eqnarray}
\begin{array}{ll}
    [ a_i, \; a^+_j ] = \delta_{ij},   &
    [ a_i, \; a_j ] = [ a^+_i, \; a^+_j ] = 0;  \cr
    [ \hat{n}_i, \mbox{} a^+_j ] = a^+_j \delta_{ij} ,  &
    [ \hat{n}_i, \mbox{} a_j ] = - a_j \delta_{ij},
\end{array}
\label{an-cr}
\end{eqnarray}
and satisfy the following four equations to be used later
\begin{eqnarray}
\begin{array}{l}
a_i^+ |... ,m_i , ...  \rangle
    = |...,m_i+ 1,... \rangle, \cr
a_i |... ,m_i , ... \rangle
    = m_i |...,m_i - 1,... \rangle,    \cr
\hat{n}_i |... ,m_i , ... \rangle
    = m_i |...,m_i , ... \rangle, \cr
e^{a_i} |... ,m_i , ...  \rangle
    = \sum\limits_{k=0}^{m_i} C_{m_i}^k |...,k,... \rangle.
\label{b-equation}
\end{array}
\end{eqnarray}

\par
Thus, the corresponding Fock representation of ${\cal SU}_n(2)$
may be obtained directly from the indecomposable representation $\rho$ as
\begin{equation}
\begin{array}{rl}
    F(J_3) |m_1,m_2,m_3 \rangle = & |m_1,m_2,m_3+1 \rangle
        + (m_1 - m_2) |m_1,m_2,m_3 \rangle, \\
    F(J_+) |m_1,m_2,m_3 \rangle = & |m_1+1,m_2,m_3 \rangle, \\
    F(J_-) |m_1,m_2,m_3 \rangle = & |m_1,m_2+1,m_3 \rangle
        -\sum\limits_{h=0}^{n} \sum\limits_{i=h}^{n}
        \sum\limits_{j=0}^{m_1-1} C_i C_{i}^h   \\
    {} & \times (j-m_2)^{i-h} |m_1-1, m_2, m_3+ h \rangle.
\label{rep-fock}
\end{array}
\end{equation}
In terms of Eq. (\ref{b-equation}), we can obtain from 
Eq. (\ref{rep-fock}) by using the induction the inhomogeneous 
three-boson realization of ${\cal SU}_n(2)$
\begin{equation}
\begin{array}{rl}
B(J_3) = & a_3^+ + \hat{n}_1 - \hat{n}_2, \\
B(J_+) = & a_1^+, \\
B(J_-) = & a_2^+ + \sum\limits_{h=0}^{n}
    \sum\limits_{i=h}^{n} (-)^{i-h+1}
    C_i  C_{i}^h (a_3^+)^h a_1          \\
{} &    \times \left[ \frac{1}{i-h+1} \sum\limits_{q=0}^{i-h}
    \hat{n}_2^{i-h-q} \left( \hat{n}_2 - \hat{n}_1 \right)^q \right.
+ \frac{1}{2} \sum\limits_{q=0}^{i-h-1} \hat{n}_2^{i-h-1-q}
    ( \hat{n}_2 - \hat{n}_1 )^q \\
{} &  \left. +\sum\limits_{r=0}^{[n/2]-1} D_r
\prod\limits_{p=0}^{2r}
      ( i- h- p ) \sum\limits_{q=0}^{i-h-2(r+1)}
    \hat{n}_2^{i-h-2(r+1)-q}  ( \hat{n}_2 - \hat{n}_1 )^q
      \right],
\label{rep-boson}
\end{array}
\end{equation}
where the symbol $[x]$ means taking integer of $x$, and $D_r$'s are real
constant numbers with respect to $n$, 
the first five constant numbers obtained by using software
``MATHEMATICA" are $D_0 = 1/ 12$, $D_1 = - 1/ 720$, $D_2 = -1/
30240$, $D_3 = -1/ 1209600$, $D_4 = 1/ 47900160$,.... 
We notice that only $a_3^+$ appears in Eq. (\ref{rep-boson}), 
whereas its adjoint $a_3$ does not, hence, if we replace $a_3^+$ in Eq. 
(\ref{rep-boson}) with an arbitrary function of $a_3^+$ and $a_3$, 
then the new three-boson realization obeys the commutation relations 
(\ref{n-am}) of ${\cal SU}_n(2)$ as well. 
When $C_1=2$ and $C_0= C_j =0$ ($j>1$), Eq. (\ref{rep-boson}) 
becomes the results of SU(2), which are the Hermitian conjugate of the
results of Ref. \cite{dgl,fs}. 

\par
In order to realize ${\cal SU}_n(2)$ by using the less bosons (one or two sets),
we may further consider the quotient spaces $U({\cal SU}_n(2))/I_i$ of 
$U({\cal SU}_n(2))$, and obtain, by the same approach, the other inhomogeneous
boson realizations of ${\cal SU}_n(2)$ from the induced representations on 
$U({\cal SU}_n(2))/I_i$, whose acting spaces are the subspaces of the Fock space
(\ref{fock-space}).

\par
(1) On the quotient space $U({\cal SU}_n(2))/I_1$, where the left
ideal $I_1$ is generated by one element $J_3 - \Lambda 1$
($\Lambda$ is a complex number), with basis
\begin{equation}
    \{ X(m_1,m_2) \equiv X(m_1,m_2,0)\mbox{mod} I_1 |m_1,m_2 \in \aleph \},
\end{equation}
the indecomposable representation $\rho$, given by Eq. (\ref{indecom-rep}), 
induces a representation
\begin{equation}
\begin{array}{rl}
\rho_1(J_3) X(m_1,m_2) = & (\Lambda+ m_1- m_2) X(m_1,m_2), \\
\rho_1(J_+) X(m_1,m_2) = & X(m_1+1,m_2), \\
\rho_1(J_-) X(m_1,m_2) = & X(m_1,m_2+1)-\sum\limits_{h=0}^{n}
     \sum\limits_{i=h}^{n} \sum\limits_{j=0}^{m_1-1}
     C_i C_{i}^h \\
{} & \times (j-m_2)^{i-h}\Lambda^h X(m_1-1,m_2). \label{1rep}
\end{array}
\end{equation}
In the process of calculating Eq. (\ref{1rep}), the property $
\rho_1 (J_3) 1 = \Lambda 1$ has been utilized. The operator
$\rho_1(J_3 )$ on $U({\cal SU}_n(2))/I_1$ has the eigenvector
$X(m_1,m_2)$ corresponding to the eigenvalue $\Lambda + m_1 - m_2$. 
Thus, by making use of Eq. (\ref{b-equation}),
we may obtain from the Fock representation that corresponds to 
Eq. (\ref{1rep}) the inhomogeneous two-boson realization of 
${\cal SU}_n(2)$
\begin{equation}
\begin{array}{rl}
B_1(J_3) = & \Lambda + \hat{n}_1- \hat{n}_2,  \\
B_1(J_+) = & a_1^+, \\
B_1(J_-) = & a_2^+ + \sum\limits_{h=0}^{n}
    \sum\limits_{i=h}^{n} (-)^{i-h+1}
    C_i C_{i}^h \Lambda^h a_1             \\
{} & \times \left[ \frac{1}{i-h+1}
      \sum\limits_{q=0}^{i-h} \hat{n}_2^{i-h-q}
    ( \hat{n}_2- \hat{n}_1 )^q + \frac{1}{2} \sum\limits_{q=0}^{i-h-1}
    \hat{n}_2^{i-h-1-q} (\hat{n}_2- \hat{n}_1 )^q \right. \\
{} & \left. +\sum\limits_{r=0}^{[n/2]-1}
     D_r \prod\limits_{p=0}^{2r} ( i- h- p )
     \sum\limits_{q=0}^{i-h-2(r+1)} \hat{n}_2^{i-h-2(r+1)-q}
    (\hat{n}_2- \hat{n}_1 )^q
   \right].
\label{1rep-boson}
\end{array}
\end{equation}
The above result may also be obtained from Eq. (\ref{rep-boson}) by replacing
$a^+_3$ with $\Lambda$.

(2) On the quotient space $U({\cal SU}_n(2))/I_2$, where the left ideal $I_2$ 
is generated by one element $J_- - \lambda 1$ ($\lambda$ is a complex number), 
with basis
\begin{equation}
    \{ X(m_1,m_3) \equiv X(m_1,0,m_3) \mbox{mod} I_2 |m_1, m_3 \in \aleph \},
\end{equation}
the indecomposable representation $\rho$, Eq. (\ref{indecom-rep}), induces
a representation
\begin{equation}
\begin{array}{rl}
\rho_2(J_3) X(m_1,m_3) = & X(m_1, m_3+1)+ m_1 X(m_1, m_3), \\
\rho_2(J_+) X(m_1,m_3) = & X(m_1+1, m_3), \\
\rho_2(J_-) X(m_1,m_3) = & \lambda\sum\limits_{r=0}^{m_3}
    C_{m_3}^r X(m_1,r) \\
{} & -\sum\limits_{h=0}^{n} \sum\limits_{i=h}^{n}
    \sum\limits_{j=0}^{m_1-1} C_i C_{i}^h
      j^{i-h} X(m_1-1, m_3+ h),
\label{2rep}
\end{array}
\end{equation}
by means of the property $ \rho_2 (J_-) 1 = \lambda 1$.
With the help of Eq. (\ref{b-equation}), solving the Fock representation 
that corresponds to Eq. (\ref{2rep}) gives rise to
\begin{equation}
\begin{array}{rl}
B_2(J_3) = & a_2^+ + \hat{n}_1, \\
B_2(J_+) = & a_1^+, \\
B_2(J_-) = & \lambda e^{a_2}+ \sum\limits_{h=0}^{n}
  \sum\limits_{i=h}^{n} (-)^{i-h+1}
    C_i C_{i}^h a_1                    \\
{} & \times \left[ \frac{1}{i-h+1} (-\hat{n}_1)^{i-h}
      + \frac{1}{2}(-\hat{n}_1)^{i-h-1} \right.  \\
{} & \left.
   + \sum\limits_{r=0}^{[n/2]-1} D_r \prod\limits_{p=0}^{2r}
    ( i-h-p ) (-\hat{n}_1)^{i-h-2(r+1)}
   \right] (a_2^+)^h.
\label{2rep-boson}
\end{array}
\end{equation}
Compared with Eq. (\ref{1rep-boson}), clearly, Eq. (\ref{2rep-boson}) 
gives another kind of two-boson realization of ${\cal SU}_n(2)$.

(3) On the quotient space $U({\cal SU}_n(2))/I_3$, where the left
ideal $I_3$ is generated by two elements \{$J_-$, $J_3 - \kappa
1$ \} ($\kappa$ is a complex number), with basis
\begin{equation}
    \{ X(m_1) \equiv X(m_100)\mbox{mod} I_3 |m_1 \in \aleph \},
\end{equation}
the indecomposable representation $\rho$, Eq. (\ref{indecom-rep}), induces a 
representation
\begin{equation}
\begin{array}{rl}
\rho_3(J_3) X(m_1) = & (\kappa+ m_1) X(m_1),  \\
\rho_3(J_+) X(m_1) = & X(m_1+1),  \\
\rho_3(J_-) X(m_1) = & -\sum\limits_{h=0}^{n}
    \sum\limits_{i=h}^{n} \sum\limits_{j=0}^{m_1-1} C_{i}^h
      j^{i-h} \kappa^h X(m_1-1),
\label{3rep}
\end{array}
\end{equation}
in virtue of the property $ \rho_3 (J_3) 1 = \kappa 1$. The
operator $\rho(J_3 )$ on $U({\cal SU}_n(2))/I_3$ has the
eigenvector $X(m_1)$ with $\kappa + m_1$ as the eigenvalue. From
the Fock representation that corresponds to (\ref{3rep}), we may
obtain by making use of Eq. (\ref{b-equation})
the inhomogeneous single-boson realization of ${\cal SU}_n(2)$
\begin{equation}
\begin{array}{rl}
B_3(J_3) = & \kappa + \hat{n}_1, \\
B_3(J_+) = & a_1^+, \\
B_3(J_-) = & -\sum\limits_{h=0}^{n}
  \sum\limits_{i=h}^{n} C_i C_{i}^h \kappa^h a_1
    \left[ \frac{1}{i-h+1} \hat{n}_1^{i-h}
      - \frac{1}{2} \hat{n}_1^{i-h-1} \right. \\
{} & \left. + \sum\limits_{r=0}^{[n/2]-1}
    D_r \prod\limits_{p=0}^{2r} ( i-h-p )
    \hat{n}_1^{i-h-2(r+1)}
   \right].
\label{3rep-boson}
\end{array}
\end{equation}
When $C_1=2$ and $C_0= C_j =0$ ($j>1$), combined with $\kappa =
-j$, Eq. (\ref{3rep-boson}) becomes the so-called Gel'fand-Dyson
representation of SU(2). \cite{dyson}

In summary, we have obtained the explicit expressions for 
the inhomogeneous single-, two- and three-boson realizations of 
${\cal SU}_n(2)$ for arbitrary $n$ by solving the Fock representaions 
of ${\cal SU}_n(2)$ that correspond to the indecomposable representation
on the space of universal enveloping algebra $U({\cal SU}_n(2))$ and
to the induced representations on the quotient spaces $U({\cal SU}_n(2))/I_i$. 
Clearly, These realizations include the results of ${\cal SU}_2(2)$ \cite{rjr}
as a special case with $C_i = 0$ ($i>2$).
For the universal enveloping algebra $U({\cal SU}_n(2))$, 
we may also choose the other bases by ordering three generators 
$J_+$, $J_-$, $J_3$ in different sequence, on which the corresponding boson 
realizations may be related by symmetry considerations. 
Due to the tight relations between boson operators and differential 
operators, \cite{bl} for a simple example, 
$a_i$ $\leftrightarrow$ $\partial / \partial {x_i}$ and
$a^+_i$ $\leftrightarrow$ $x_i$,
the inhomogeneous differential realizations of ${\cal SU}_n(2)$ may 
be obtained directly from their inhomogeneous boson realizations.
The results obtained in this paper will be applied to the potential group 
approach \cite{agi} with respect to ${\cal SU}_n(2)$ to study for some typical 
quantum mechanical systems the corresponding bound and scattering states, 
(nonlinear) coherent states and so on.

\section*{Acknowledgments}
This work is supported by National Natural Science Foundation of China
(19905005), Major State Basic Research Development Program (G2000077400)
and Tsinghua Natural Science Foundation (985 Program).

\hspace{1cm}

\end{document}